\documentclass{article}
\pdfoutput=1
\usepackage{amsmath,amsfonts}
\usepackage{algorithmic}
\usepackage{array}
\usepackage[caption=false,font=normalsize,labelfont=sf,textfont=sf]{subfig}
\usepackage{textcomp}
\usepackage{stfloats}
\usepackage{url}
\usepackage{verbatim}
\usepackage{graphicx}
\def\BibTeX{{\rm B\kern-.05em{\sc i\kern-.025em b}\kern-.08em
    T\kern-.1667em\lower.7ex\hbox{E}\kern-.125emX}}
\usepackage{balance}
\begin{document}
\title{Fast Discrete Fourier Transform algorithms requiring less than 0($N\log N$) multiplications}
\author{Ryszard M. Stasi\'{n}ski \\
Institute of Multimedia Telecommunications \\
Poznan University of Technology, Pozna\'{n}, Poland \\
	e-mail: {\tt ryszard.stasinski@put.poznan.pl}\\
	}

\markboth{}{Almost multiplierless DFT algorithms}


\maketitle

\begin{abstract}
In the paper it is shown that there exist infinite classes of fast DFT algorithms 
having multiplicative 
complexity lower than $O(N\log N)$, i.e. smaller than their arithmetical complexity. The derivation starts with 
nesting of Discrete Fourier Transform (DFT) of size $N = q_1\cdot q_2\cdot \ldots q_r$, where $q_i$ are powers of 
prime numbers: DFT is mapped into multidimensional one, 
Rader convolutions of $q_i$-point DFTs extracted, and combined into multidimensional convolutions 
processing data in parallel. Crucial to further optimization 
is the observation that multiplicative complexity of such algorithm is upper bounded by $0(N\log M_{max})$, where $M_{max}$ 
is the size of the greatest structure containing multiplications. 
Then the size of the structures is diminished: Firstly, computation of a circular convolution 
can be 
done as in Rader-Winograd algorithms. Secondly, multidimensional convolutions can be computed using polynomial transforms.
It is shown that careful choice of $q_i$ values leads to important reduction of 
$M_{max}$ value:
Multiplicative complexity of 
the new DFT algorithms is $O(N\log ^c \log N)$ for $c\le 1$, while for more addition-orietnted ones 
it is $O(N\log ^{1/m} N)$, $m$ is a natural number denoting class of $q_i$ values. 
Smaller values of $c$, $1/m$ 
are obtained for algorithms requiring more additions, part of algorithms for $c = 1$, $m=2$ have 
arithmetical complexity 
smaller than that for the radix-2 FFT for any comparable DFT size, 
and even lower than that of split-radix FFT for $N\le 65520$. 
The approach can be used for finding theoretical lower 
limit on the DFT multiplicative complexity.
\end{abstract}

\begin{quote}
\end{quote}

\section{Introduction}
\label{intro}

\noindent The re-discovery of Fast Fourier Transform 
(FFT) in 1965 \cite{fft65} resulted in a wave of related publications, e.g. 
there were two special issues on this subject of IEEE Transactions on Audio 
Electroacoustics in 1967, and 1969 \cite{au6769}. It also raised some fundamental 
questions, is the computational complexity of $N$-point DFT $O(N\log N)$, 
or lower \cite{aho}? Or at least, what is the lower bound on DFT multiplicative 
complexity in the field of complex numbers? The answer to the second question is provided by Winograd 
in \cite{wino80}, 
it is $O(N)$, but the result is obtained for excessive number of 
additions, surpassing even $O(N^2)$, the complexity of DFT definition formula.
More influental was his other work \cite{wino}, which triggerd the second wave 
of publications, 
including those on the Winograd-Fourier transform algorithm (WFTA), which 
realization was presented in \cite{silver}, and with some improvements in \cite{burrb}. 
At the same time important results were obtained 
for multidimensional DFT 
computation by Nussbaumer \cite{nuss}, \cite{nussq}, who devised polynomial 
transforms, finally split-radix FFT was introduced \cite{srfft}. 
Winograd's works raised the question: to what extent it is reasonable to reduce 
the number of multiplications needed for computation of DFT, while keeping
its computational cost low? 

Simple close form formulae on the number of operations (\ref{fftma}) cause that split-radix FFT \cite{srfft} 
is a convenient yardstick for eveluating performance of DFT algorithms, but this is not the best FFT. 
Firstly, in \cite{optfft} its version with reduced number of additions was presented. Recently it was 
shown in \cite{stas22} that there are several algorithms having smaller arithmetical, or at least multiplicative 
complexity than split-radix FFT. The smallest arithmetical complexity has radix-6 FFT from \cite{martens}, the 
used in section \ref{mults} radix-6 FFT from \cite{stas94} has exceptionally small number of multiplications.

Recently the problem of DFT computational complexity is attacked from another angle,
additions and multiplications are not counted separately, only the $N$-point DFT computation 
is transformed into integer multiplication problem of size $bN$ \cite{harvey}, 
$N$ is the sequence length, $b$ is the integer representation length. 
Then in the next paper \cite{harvey19} it is proved that it exists
an integer multiplication algorithm requiring $O(b\log b)$ bit operations, which means that DFT bit 
complexity is $O(bN\log bN)$. 
The transformation is based on paper \cite{filterdft}, where DFT 
computation is done by calculation of a convolution.
Note that if bit complexity of an addition is $O(b)$, 
then bit complexity of $O(N\log N)$ additions is $O(bN\log bN)$, see (\ref{M2A}), (\ref{Acond}).

This paper is going the way pointed out by Winograd \cite{wino80}, this time 
transformation of DFT computation problem into that of convolution calculation is 
based on a generalization of Rader idea \cite{11}. 
There is no assumption concerning the way in which real multiplications and additions are done.
It is shown that there are DFT algorithm classes, for which
\begin{equation}
	\label{M2A}
	\frac{M(N)}{A(N)} \longrightarrow{0} \quad {\rm for} \quad N \rightarrow \infty
\end{equation}
where $M(N)$, $A(N)$ are multiplication and addition numbers, and 
\begin{equation}
	\label{Acond}
	A(N) = O(N\log N).
\end{equation}
The result is obtained using the technique from \cite{stas87}.
Algorithm construction starts with nesting of Rader-Winograd DFTs, section \ref{idea}, 
that can be applied to any DFT of size: 
\begin{equation}
	\label{anyn}
	N = p_1^{k_1}\cdot p_2^{k_2}\cdot p_3^{k_3}\cdot \ldots
\end{equation}
$p_i$ are prime numbers. Whenever at least two numbers  
$p_i - 1$, $p_j - 1$ have common factors, polynomial transforms 
can be used for reducing multiplicative complexity of the algorithm, 
section \ref{optimal}. 
Before \cite{stas87} polynomial transforms were used only for optimization of 
$63 = 9\cdot 7 $ and $80 = 16\cdot 5$ 
DFT modules \cite{nuss81}. 
For full development of the algorithm 
Winograd's findings on DFT computation are necessary, section \ref{winorader}, as well as for 
calculation of lower bound on the number of 
multiplications for such $N$, section \ref{dftalg}, Theorem 1. 

To meet condition (\ref{M2A}) majority of 
primes in (\ref{anyn}) should have possibly large common factors 
for $p_i - 1$. Condition (\ref{Acond}) implies the use of FFT for 
polynomial transforms of possibly small radices, section \ref{adds}. 
Lack of this restriction in \cite{stas87} caused that the number of additions grew 
faster than $O(N\log N) $.
As a result 
primes are searched in specific families of numbers (\ref{primes23}). 
The assumption concerning density of these primes, hence, the validity of proven in appendix A Theorems 2, and 3 
seems to be supported by experimental data, 
section \ref{mults}. 
Operation numbers for the best 
algorithms for $N \le 65520$ are provided in appendix \ref{reala}.   

\section{Nesting}
\label{idea}

According to Rader \cite{11} DFT computation when its size $N$ is a 
prime number can be done by an ($N-1$)-point circular convolution:
\begin{equation}
	\label{Rader}
	X(k) \Longleftarrow \left\{ \begin{array}{l}
		X(0) = x'(0) \\                                          
		X(a_{k}) = \sum_{n=0}^{N-2} x'(a_{n})W_{N}^{a_{k-n}}
	\end{array}
	\right.
\end{equation}
where:
\begin{equation}
	\label{dft2conv}
	x'(n) = \left\{ \begin{array}{ll}
		\sum_{m=0}^{N-1} x(m) & \mbox{for } n=0 \\
		x(N-n) - x(0)         & \mbox{for } n\neq 0
	\end{array}
	\right.
\end{equation}
and $a_n$ are cyclic group elements modulo $N$, the group size is $N-1$. Rader algorithms exist also 
when $N$ is a power of a prime number 
\cite{stas86}.  
Their structure is more complicated, hence, a good idea is to add such DFTs to the structure of the 
developed here algorithms later, when Winograds improvements to the Rader algorithm are implemented,
section \ref{winorader}. 

The Rader algorithm can be 
represented in matrix form as follows:
\begin{equation}
	\label{rawidft}
	{\bf T} = {\bf E}\cdot {\bf H}\cdot {\bf D}
\end{equation}
where here matrix ${\bf D}$ represents operation from (\ref{dft2conv}), multiplication by matrix 
${\bf H}$ is equivalent to computations in (\ref{Rader}), while matrix ${\bf E}$ defines final inverse 
permutations. 
Note that multipliers 
are present only in two blocks of matrix ${\bf H}$: $N-1$-point one realizing convolution, and another one 
of size 1 needed to transfer sample $x'(0)$ to the algorithm output (\ref{Rader}). 

When $N$ is a product of prime numbers $q_i$, $i=1, 2,\ldots r$, the $q_i-1$-point circular convolutions can be nested. 
The technique is applicable for any 
$N = N_1\cdot N_2\cdot \ldots N_r$ when each $N_i$ is 
mutually prime with any other $N_j$. DFTs of such size can be mapped into 
$r$-dimensional ones of size $N_1\times N_2\times \ldots N_r$ by 
the following permutation of input sequences: 
\begin{equation}
	\label{m_Dt}
	x[(\frac{N}{N_{1}}n_{1}+\frac{N}{N_{2}}n_{2}+ \ldots 
	+\frac{N}{N_{r}}n_{r}) \mbox{ mod } N] 
	\Longrightarrow x(n_{1},n_{2},\ldots , n_{D}).
\end{equation}
In this way prime factor FFT algorithms (PFA FFT) are obtained 
\cite{au6769}. 
The mapping is reversible, $r$-dimensional DFT for such dimensions can be mapped into one-dimensional one. 
In the paper both mappings are extensively used. 
In matrix form $r$-dimensional DFT can be represented 
as follows:
\begin{equation}
	\label{rowcol}
	{\sf T} = {\bf T_{\rm r}\otimes \ldots \otimes T_{\rm 2}\otimes T_{\rm 1}}, 
\end{equation}
where $\otimes$ is the Kronecker product symbol, and ${\bf T}_{\rm i}$ is one dimensional $N_i$-point  
DFT matrix. Then: 
\begin{equation}
	\label{EHD}
	\begin{array}{l}
		{\sf T} = 
		{\bf (E_{\rm r}\cdot H_{\rm r}\cdot D_{\rm r}) \otimes \ldots \otimes 
			(E_{\rm 2}\cdot H_{\rm 2}\cdot D_{\rm 2}) \otimes 
			(E_{\rm 1}\cdot H_{\rm 1}\cdot D_{\rm 1}) = } \\
		{\bf = (E_{\rm 1}\otimes E_{\rm 2}\otimes \ldots \otimes E_{\rm r})
			(H_{\rm r}\otimes \ldots \otimes H_{\rm 2}\otimes H_{\rm 1})
			(D_{\rm r}\otimes \ldots \otimes D_{\rm 2}\otimes D_{\rm 1})} = \\
		= {\sf  E\cdot H\cdot D}
	\end{array}
\end{equation}
matrices ${\bf H}_{\rm i}$ are {\it nested} inside the structure, matrices 
${\sf T}$, ${\sf E}$, ${\sf H}$, and ${\sf D}$ are $r$-dimensional ones. Matrix ${\sf H}$ 
is a block-diagonal one, its blocks are  defined by Kronecker products of blocks of ${\bf H_{\rm i}}$ 
matrices. Nested is possible, as linear multi-dimensional transforms are commutable \cite{nuss81}. 
The commutability 
is used few times in this paper for algorithm optimization. 

Assume now that $N_i = q_i$. The blocks of ${\bf H_{\rm i}}$ are obtained from blocks of individual one-dimensional DFTs: 
from $1\times 1\times \ldots 1$-point one ($r$ times repeated 1), to the 
$(q_1-1)\times (q_2-1)\times \ldots (q_r-1)$-point one, and are of size from 1 to 
$M_{max}=(q_1-1)\cdot (q_2-1)\cdot \ldots (q_r-1)$. The blocks 
represent $r$-dimensional circular convolutions. 
The fundamental 
properties of the structure (\ref{EHD}) are:
\begin{itemize}
	\item Input data vector is split into non-overlapping segments independently processed by blocks of matrix ${\sf H}$.
	\item Multipliers are present only in these blocks.
	\item The sum of block sizes is $N$.
	\item Computational complexity of vector multiplication by each block is upper bounded by 
	$0(M_i\log M_i)$, where $M_i$ is the block size.
	\item As a consequence multiplicative complexity of the structure (\ref{EHD}) is upper bounded by 
	$0(N\log M_{max})$, where $M_{max}$ is the greatest block size. 
\end{itemize}
These properties are kept in further derivation of the algorithms.

\section{Winograds improvements to Rader algorithm}
\label{winorader}

Coefficients of a product of polynomials $Y(Z) = H(Z)X(Z)$ can be obtained by 
convolving coefficients of polynomials $H(Z)$, and $X(Z)$. If the result is reduced modulo $Z^M - 1$, the $M$-point 
circular convolution of coefficients is obtained. 
Let us introduce based on the Chinese Remainder Theorem approach to 
derivation of polynomial product algorithms in rings of residues 
\cite{nuss81}:
\begin{quote}
	{\bf Algorithm 1:}
	\nopagebreak
	
	Task: Compute a polynomial product $Y(Z) = H(Z)\cdot X(Z)$ modulo $Q(Z)$. Polynomial $Q(Z)$ 
	can be factored as follows:
	\[
	Q(Z) = \prod_{i} Q_i(Z)
	\]
	1. Compute residues modulo all polynomials $Q_i(Z)$ (stage of polynomial reductions):
	\[
	X_i(Z) = X(Z) \, {\rm mod}\, Q_i(Z),   H_i(Z) = H(Z)\, {\rm mod}\, Q_i(Z)
	\]
	2. Multiply the residues:
	\[
	Y_i(Z) = H_i(Z)\cdot X_i(Z)\, {\rm mod}\, Q_i(Z)
	\]
	3. Reconstruct $Y(Z)$ from its residues $Y_i(Z)$ (stage of polynomial reconstruction).
\end{quote}
Assume that steps 1. and 3. can be done without multiplications. Then, the matrices ${\bf D}$, 
${\bf H}$, and ${\bf E}$ in (\ref{rawidft}) can be redefined: 
matrix ${\bf D}$ represents operation from (\ref{dft2conv}) followed by a stage of 
polynomial reductions of Algorithm 1, matrix ${\bf E}$ represents polynomial reconstruction, 
and multiplications by blocks of matrix ${\bf H}$ do polynomial products modulo 
divisors of $Q(Z) = Z^M -1$. Note that multiplications are associated only with multiplication 
by the ${\bf H}$ matrix.

In the Rader-Winograd DFT algorithms polynomials $Q_i(Z)$ of the smallest possible rank 
are used, i.e. irreducible divisors of $Z^M -1$, called 
cyclotomic polynomials. In this paper notation for them is 
$P_N(Z)$, where $N$ is the smallest number $K$, for which the cyclotomic polynomial 
$P_N(Z)$ is a divisor of the polynomial $Z^K - 1$. 
Sum of ranks of divisors of  
$Z^M -1$ is $M$, which means that modified matrices in (\ref{rawidft}) are still square and of size $N$. 

Cyclic groups exist for any power of a prime number, and Winograd proposed efficient algorithms 
for few powers of primes: $N = 4, 8, 9, 16$. 
The full description of Rader-type DFT algorithms for powers of primes is 
given in \cite{stas86}, power of number 2 algorithms are also outlined in \cite{duh84}. 
What is important here, they also can be represented 
by formula (\ref{rawidft}) with multiplication operations appearing only in the matrix ${\bf H}$, 
having size $N$, too. 
When including such DFTs in the structure (\ref{EHD}) it is worth noting that for $N=p^r$, $p$ is an odd prime, polynomial products
appearing in blocks of ${\bf H}$ are modulo irreducibile divisors of $Z^{p^s}-1$, $s<r$, and of 
$Z^{p-1}-1$. 
For $N$ being a power of number 2 there are pairs of polynomial products 
modulo $P_{2^{s}}(Z)$, $s=0,1,\ldots ,r-2$, $r>1$. 
 
\section{Optimization of multidimensional polynomial products}
\label{optimal}

\subsection{Basic algorithms \cite{nuss81}}
\label{convolutions}

The general structure of the most frequently used class of polynomial product algorithms is provided in Algorithm 1. 
Let us consider it's special case for (divisors of) $Z^N-1$:
\begin{quote}
	{\bf Algorithm 1F:}
	\nopagebreak	
	
	Task: Compute the polynomial product $Y(Z) = H(Z)\cdot X(Z)$ modulo $Z^N-1$, or its divisor. Such polynomial 
	can be factored as follows:
	\[
	Q(Z) =  \prod_{k} (Z - W_N^k), \quad 0 \le k < N, 
	\] 
	$W_N = \exp (-j2\pi /N)$. For $P_N(Z)$ all $k$ mutually prime with $N$ are used.
	
	Steps 1-3 are the same as in Algorithm 1.
\end{quote}

Residue modulo a binomial is a scalar:
\[
X(Z)\, {\rm mod}\, (Z - W_N^k) = \sum_{n=0}^{M-1} x(n) W_N^{kn} 
= \sum_{n=0}^{M-1} x(n) \exp (-j2\pi kn/N) = X(k)
\] 
As can be seen, for $P_N(Z)$ the residues are computed using a "cropped" DFT formula, DFTs for 
$k$ mutually prime with $N$ are called reduced $N$-point DFTs \cite{nuss81}. 
Effective algorithms for them can be easily extracted from "full" ones. 
Step 3. in Algorithm 1F is done using the inverse (reduced) DFT. Note that actual size of reduced
$N$-point DFT is $M$, the rank of polynomial $P_N(Z)$.

If a reduced DFT algorithm is not sufficiently effective, a polynomial product can be 
computed by a "brute force" approach:
\begin{quote}
	{\bf Algorithm 2:}
	
	Task: Compute the polynomial product $Y(Z) = H(Z)\cdot X(Z)$ modulo $P_N(Z)$. 
	
	1. Compute fast convolution algorithm of coefficients of $H(Z)$ and $X(Z)$.  
	
	2. Form a polynomial from the result, and then reduce it modulo $P_N(Z)$.
	
	Comment: An obvious choice in step 1 is Algorithm 1F using FFT of size not 
	smaller than $2M-1$. 
\end{quote}

\subsection{Multidimensional polynomial products}
\label{multiD} 

Let us exploit circular shift property for reducing 
multiplicative complexity of the $N\times N$-point DFT:
\begin{equation}
	\label{shiftDFT}
	\begin{array}{l}\vspace{.1in}
	\sum_{n_2 =0}^{N-1} [\sum_{n_1 =0}^{N-1} x(n_1 , (n_2 - k_1 n_1)_N )] W_N ^{k_2\cdot n_2} = \\
	\vspace{.1in}
	= \sum_{n_1 =0}^{N-1} \sum_{n_2 =0}^{N-1} x(n_1 , (n_2 - k_1 n_1)_N ) W_N ^{k_2\cdot n_2} = \\
	\vspace{.1in}
	= \sum_{n_1 =0}^{N-1} W_N ^{k_2 k_1 n_1 } \sum_{n_2 =0}^{N-1} x(n_1 , n_2 ) W_N ^{k_2\cdot n_2} = \\
	= X((k_2 k_1)_N ,k_2)
\end{array}
\end{equation}
$(.)_N$ denotes reduction modulo $N$. $(k_2 k_1)_N$ takes on all $k_1$ values only when 
$k_2$ is mutually prime witn $N$, hence, in (\ref{shiftDFT}) some DFT samples 
$X(k_1, k_2 )$ are not computed, while some other more than once.  
Taking into account multiplierless additions in the inner formula of (\ref{shiftDFT}) a DFT in the ring of 
residue polynomials modulo $Z^N -1$ can be introduced:
\begin{equation}
	\label{ringpt}
	X_{k_1} (Z) = \sum_{n_1 =0}^{N-1} x_{n_1} (Z) Z^{k_1 n_1} {\rm mod} (Z^N -1)
\end{equation}
where coefficients of residue polynomials $x_{n_1} (Z)$ are $x(n_1 , n_2 )$.
Indeed,
polynomial $Z$ is here the primitive root of unity of rank $N$:
\[
Z^N {\rm mod} (Z^N -1) = 1, \quad Z^n {\rm mod} (Z^N -1) \ne 1 \quad {\rm if} \quad n<N.
\]
For $N_1 \times N_2$ DFT when $N_1 $ is a 
divisor of $N_2 $ the primitive root of unity in (\ref{ringpt}) is simply $Z^{N_2 /N_1 }$.

To avoid problematic calculations in (\ref{shiftDFT}) 
polynomials $x_{n_1} (Z)$ are reduced modulo $P_N (Z)$, in this way polynomial transform is formulated 
\cite{nuss}, \cite{nussq}, \cite{nuss81}:
\begin{equation}
	\label{pt}
	X_{k_1} (Z) = \sum_{n_1 =0}^{N-1} x_{n_1} (Z) Z^{k_1 n_1} {\rm mod} P_N (Z)
\end{equation}
Note that residue polynomials modulo $P_N(Z)$ form a field. To conclude (\ref{shiftDFT}), calculate reduced $N$-point DFTs:
\begin{equation}
	\label{redpt}
	X((k_2 k_1)_N ,k_2) = X_{k_1} (Z) {\rm mod} (Z - W_N^{k_2})
\end{equation}
for all $k_2$ mutually prime with $N$.

Let us consider two-dimensional version of Algorithm 1F for 
cyclotomic polynomials $P_{N_1}(Z_1)\times P_{N_2}(Z_2)$ when $N_1$ 
is a divisor of $N_2$. It is computed using $N_1\times N_2$-point 
reduced in both dimensions forward and inverse DFTs. As it is shown above, 
forward DFT can be also calculated by reduced polynomial transform\footnote{Polynomial transforms are intended for 
	computation of polynomial products modulo $(Z_1^{N_1}-1)\times P_{N_2}(Z_2)$ \cite{nuss81}. To exploit this  
	do not decompose $N_1$-point circular convolution into polynomial products, Algorithm 1.} for 
kernel $Z^{N_2 /N_1 }$, and $k_1$ mutually prime with $N_1$ (\ref{pt}), 
followed by reduced $N_2$-point DFTs in other dimension (\ref{redpt}). Polynomial transform is defined in a field, hence, its inverse exists, and as it is DFT, its kernel is 
\[
W_{N_1}^{-1} = Z_2^{-N_2/N_1} {\rm mod} P_{N_1}(Z_2)
\]
which means that it is calculated without multiplications, too. Then 
the following algorithm is formulated:
\begin{quote}
	{\bf Algorithm 3F:}
	
	Task: Multiplication of two-dimensional residue polynomials:
	\[
	Y(Z_1, Z_2) 
	= X(Z_1, Z_2)\cdot H(Z_1, Z_2) {\rm mod} 	P_{N_1}(Z_1)\times P_{N_2}(Z_2)
	\]
	
	when $N_1$ is a divisor of $N_2$.
	
	1. Compute $N_1$-point reduced polynomial transforms 
	of $X(Z_1, Z_2)$ 
	and $H(Z_1, Z_2)$:
	\[
	\begin{array}{l}
		X_{k_1}(Z_2) 
		= X(Z_1, Z_2)\, {\rm mod} (Z_1 - Z_2^{N_2/N_1})\, {\rm mod} P_{N_2}(Z_2), \\
		H_{k_1}(Z_2) 
		= H(Z_1, Z_2)\, {\rm mod} (Z_1 - Z_2^{N_2/N_1})\, {\rm mod}
		P_{N_2}(Z_2).
	\end{array}
	\]
	
	2. Reduce above polynomials modulo $Z_2 - W_{N_2}^{k_2}$ for $k_2$ 
	mutually prime with $N_2$ (reduced $N_2$-point DFTs).
	
	3. Multiply the resultant scalars:
	\[
	Y((k_2k_1)_N,k_2) = X((k_2k_1)_N,k_2)\cdot H((k_2k_1)_N,k_2)
	\]
	
	4. Reconstruct $Y_{k_1}(Z_2)$ (inverse reduced $N_2$-point DFTs). 
	
	5. Reconstruct $Y(Z_1, Z_2)$ (inverse reduced $N_1$-point polynomial transform).
	
\end{quote}
Note that in steps 2-4 polynomial products are computed using Algorithms 1F.
In fact any other polynomial product modulo $P_{N_2}(Z_2)$ algorithm  
can be implemented, e.g. some version of Algorithm 2. This observation leads us to 
more general version of the above algorithm:
\begin{quote}
	{\bf Algorithm 3:}
	
	Task: the same as in Algorithm 3F.
	
	Steps 1. and 3. are the same as steps 1. and 5. in Algorithm 3F.
	
	2. Compute polynomial products:
	\[
	Y_{k_1}(Z_2) = X_{k_1}(Z_2)\cdot H_{k_1}(Z_2)\, {\rm mod} P_{N_2}(Z_2)
	\]
	
\end{quote}

Algorithm 3 has the desired property: 
multiplications 
are associated only with its central stage. In the case of Winograd's Algorithm 1 reduction 
of parameter $M_{max}$ is from $N$, the convolution size, to $M$, the number of natural numbers 
smaller than $N$, and mutually prime with $N$. Algorithm 3 is much more powerfull, here reduction is 
from $M_1\cdot M_2$ to $M_2$, where $M_1$, $M_2$ are numbers of numbers mutually prime with 
$N_1$, $N_2$, respectively.

\subsection{Practical considerations}
\label{practical}

For many DFT sizes 
implementation of Algorithm 3 is not clear. Let us
consider computation of $N=56 653$-point DFT, $N=181\cdot 313$, 
neither 180 is a divisor of 312, nor 312 a divisor of 180. However, $\gcd\{180,312\}=12>1$, and 
this implies that Algorithm 3 can be implemented. To reveal this let us consider computation 
of 180- and 312-point convolutions by Algorithm 1F using PFA FFT. The algorithm is based on mapping 
(\ref{m_Dt}), hence, instead of 180- and 312-point DFTs we are dealing with $4\times 9\times 5$- and 
$8\times 3\times 13$-point ones. Then, the use of Algorithm 3 becomes clear: it should be applied 
to computation of polynomial products modulo $P_4(Z_1)\times P_8(Z_4)$ and $P_9(Z_2)\times P_3(Z_5)$.
The remaining irreducible greatest polynomial product is $P_9(Z_2)\times P_5(Z_3)\times p_8(Z_4)\times P_{13}(Z_6)$ 
having size $M_{max} = 6\cdot 4\cdot 4\cdot 12 = 1152$. Note that the same $M_{max}$ is obtained when the DFT 
size is multiplied by 61, i.e. for $N=3 455 833 = 181\cdot 313\cdot 61$. Namely, $60=4\cdot 3\cdot 5$, 
and the multipliers are divisors of 8, 9, and 5, respectively.

Let us find upper bounds on the multiplicative complexity of the above algorithms. 
In Algorithm 1F there are two FFTs, forward and inverse one, composed of small-size, hence, multiplicative-efficient 
DFT modules. Then, 
the number of multiplications of the 1152-point polynomial product can be approximated by 
$2\cdot M_{max} \log_2 M_{max}$, and as a consequence, the number of multiplication for the whole 
DFT is upper bounded by $2N\log_2 M_{max} \approx 20\cdot N$. At the same time for FFTs of similar size, i.e. for 
$N=2^{16}$ and $N=2^{22}$ it is smaller than $N \log_2 N = 16\cdot N$ and $22\cdot N$ (\ref{fftma}). It is an open 
question if $N=3 455 833$-point DFT algorithm indeed has less multiplications per input sample than the 
$N = 2^{22}$-point one, but this is definitely not a particularly efficient algorithm. In Appendix B an $N=65520$-point 
algorithm is presented, for which the number of multiplications is $4.01\cdot N$, while the number 
of arithmetical operations is slightly smaller than for $N=2^{16}$-point split-radix FFT.

\section{Arithmetical complexity}
\label{arith}

\subsection{Lower bound on DFT multiplicative complexity}
\label{dftalg}

Optimizations of DFT structure (\ref{EHD}) done in the previus sections cause that matrix ${\bf H}$ 
represents multiplication by irreducible polynomial products. At the same time its block-diagonal 
form implies that the products are done in parallel. Then, the following theorem is valid: 
\begin{quote}
	{\bf Theorem 1:} If polynomial products represented by blocks of matrix ${\bf H}$ in 
	(\ref{EHD}) are irreducible, then the lower bound on the multiplication number 
	of DFT algorithms for $N$ given by formula (\ref{anyn}) can be obtained by summing up numbers $2K_i-1$, 
	where $K_i$ are sizes of matrix ${\bf H}$ blocks.
\end{quote}
{\em Proof:} The reasoning is the same as in the paper \cite{wino80} {\em Q.E.D.}

\subsection{Additions: proposed algorithm sizes}
\label{adds}

Appearance of polynomial transforms in stages 
preceding multiplications, and their inverses in the following stages implies that in many cases 
arithmetical complexity of the algorithm shifts from the computation of polynomial products 
to that of polynomial transforms. Arithmetical complexity 
of a polynomial transform of size $p$ being a prime number is $O(p^2)$ \cite{nuss81}, which 
means that for an ill-considered choice of DFT sizes the algorithm complexity may appear to be 
greater than $O(N\log N)$, i.e. the algorithm is not fast \cite{stas87}. The solution is 
to use small-radix FFTs for polynomial transforms. In \cite{stas87} it has been shown that 
for a radix-$p$ polynomial transform, $p$ is prime: 
\begin{equation}
	\label{agrow}
	a_{\infty } = \frac{4(p^2 - 2)}{p\log _2 p} = O(\frac{p}{\log p}).
\end{equation}
where asymptotic parameter $a_{\infty }$ is:
\begin{equation}
	\label{ainf}
	\mbox{ for } N \to \infty \mbox{:    } \;  \frac{O_A (N)}{N\log _2 N} \longrightarrow a_{\infty }, 
\end{equation}
$O_A (N)$ is the arithmetical complexity of an algorithm \cite{aho}. 
A polynomial transform is 
accompanied by its inverse, hence, the number of additions of an algorithm 
composed exclusively of radix-$p$ FFTs is $2\cdot a_{\infty }N\log_2 N$, and this value can be used for 
computing the upper bound for the presented here algoritms. For radices 2, 3, 5, 7 the parameter 
$2\cdot a_{\infty }$ is 4, 5.89, 7.92, and 9.57, respectively. On the other hand, for 
base 2 FFTs this parameter grows from $a_{\infty } = 3.78$ for the algorithm from \cite{optfft}, to
$a_{\infty } = 4$ for the split-radix FFT \cite{srfft}, and
to $a_{\infty } = 5$ for the radix-2 one \cite{fft65}, while for a relatively ''inefficient'' radix-3 FFT 
$a_{\infty } \approx 6.73$ \cite{au6769}. 
This means that algorithms based entirely on polynomial transforms for powers of number 2 (for Fermat 
prime numbers) are potentially 
as efficient as 
split-radix FFT. 
Indeed, in Table I 
there are few interesting practical algorithms from this class. 

Then, classes $m$ of DFT algorithms for products of the following primes seem to be interesting:
\begin{equation}
	\label{primes23}
	\begin{array}{cc}
		m=1: & q_2 = 2^{s_{1}}	+ 1, \\
		m=2: & q_{23} = 2^{s_{1}}\cdot 3^{s_{2}} + 1, \\
		m=3: & q_{235} = 2^{s_{1}}\cdot 3^{s_{2}}\cdot 5^{s_{3}} + 1 \\
		\ldots & \ldots \ldots \ldots
	\end{array}
\end{equation}
The classes are extended as follows: DFTs for primes from family $q_2$ (Fermat prime numbers) are 
combined with DFTs for $2^{k_1}$, those from family $q_{23}$ with 
$N = 2^{k_1}\cdot 3^{k_2}$, for family $q_{235}$ DFTs for 
$N = 2^{k_1}\cdot 3^{k_2}\cdot 5^{k_3}$ are added \cite{stas86}, etc. 

The class of interesting algorithms for $q_2$ is finite and small. Then, 
algorithms for $q_{23}$ are probably the most useful ones: 
even mix of radix-2 and radix-3 FFTs for 
polynomial transforms leads to upper bound on asymptotic coefficient $a_{\infty } \approx 4.95$, which is better than for radix-2 FFT 
for DFT: $a_{\infty } = 5$. 
Indeed, up to $N=65520$ many algorithms from this class have {\em smaller} arithmetical 
complexities than the split-radix FFT, Table I.
If the number of additions is important, it is possible to bound
number $s_2$ in (\ref{primes23}) for $q_{23}$. It seems that even for $s_2 = O(1)$ there is enough primes to 
construct algorithms of sizes $N\rightarrow \infty$. 
Their multiplicative complexity might be $o(N\log N)$ \cite{stas92}, while 
$a_{\infty } = 4$, the same as for the split-radix FFT.

\subsection{Multiplications}
\label{mults}

Done in Appendix A analysis of growing disproportion between $N$, DFT size, and $M_{max}$, the 
size of the greatest irreducible polynomial product of the algorithm for primes (\ref{primes23}) leads to the following Theorem: 
\begin{quote}
	{\bf Theorem 2:} There exist fast DFT algorithms having multiplicative complexity
	\[
	O(N\log ^c \log N), c\le 1
	\] 
	i.e. lower than their arithmetical complexity $O(N\log N)$.
	
	Note: For primes 
	$q_{23}$ $c=1$. If every Fermat number were prime, multiplicative 
	complexity for $q_2$ primes would be $O(N\log^{0.5} N)$ \cite{stas92}.
\end{quote}

In this section evaluations on which the Theorem 2 is based are illustrated by results for 
real data: 
\begin{figure}
	\begin{center}
		\includegraphics{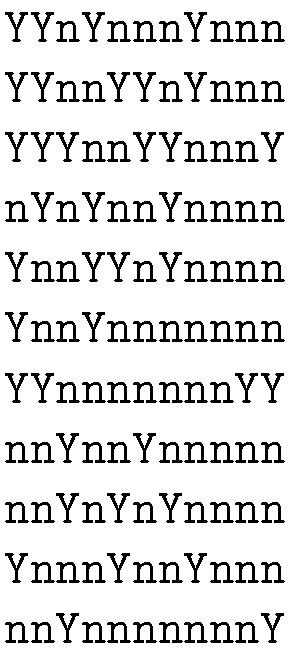}
	\end{center}
	\caption{Map 
		showing properties of numbers from which class $q_{23}$ of primes is selected 
		for $u\le 10$. On the crossing of $i$-th column and $j$-th row there is an answer if the 
		number $2^i 3^j + 1$ is prime ({\tt Y} - yes, {\tt n} - no), $i \ge 1$, $j\ge 0$.
		Answers for potential new primes for $u = 10$ 
		are on the right and bottom sides of the map, 
		there are 4 primes: 
		$2^{11}\cdot 3^2 + 1, 2^{11}\cdot 3^6 + 1, 2^{11}\cdot 3^{10} + 1$, and $2^3\cdot 3^{10} + 1$.
		In the top row letters {\tt Y} point at Fermat primes: 3, 5, 17, and 257.}
	\label{map}
\end{figure}		
\begin{figure}[h]
	\centering
	\includegraphics[height=0.4\textheight]{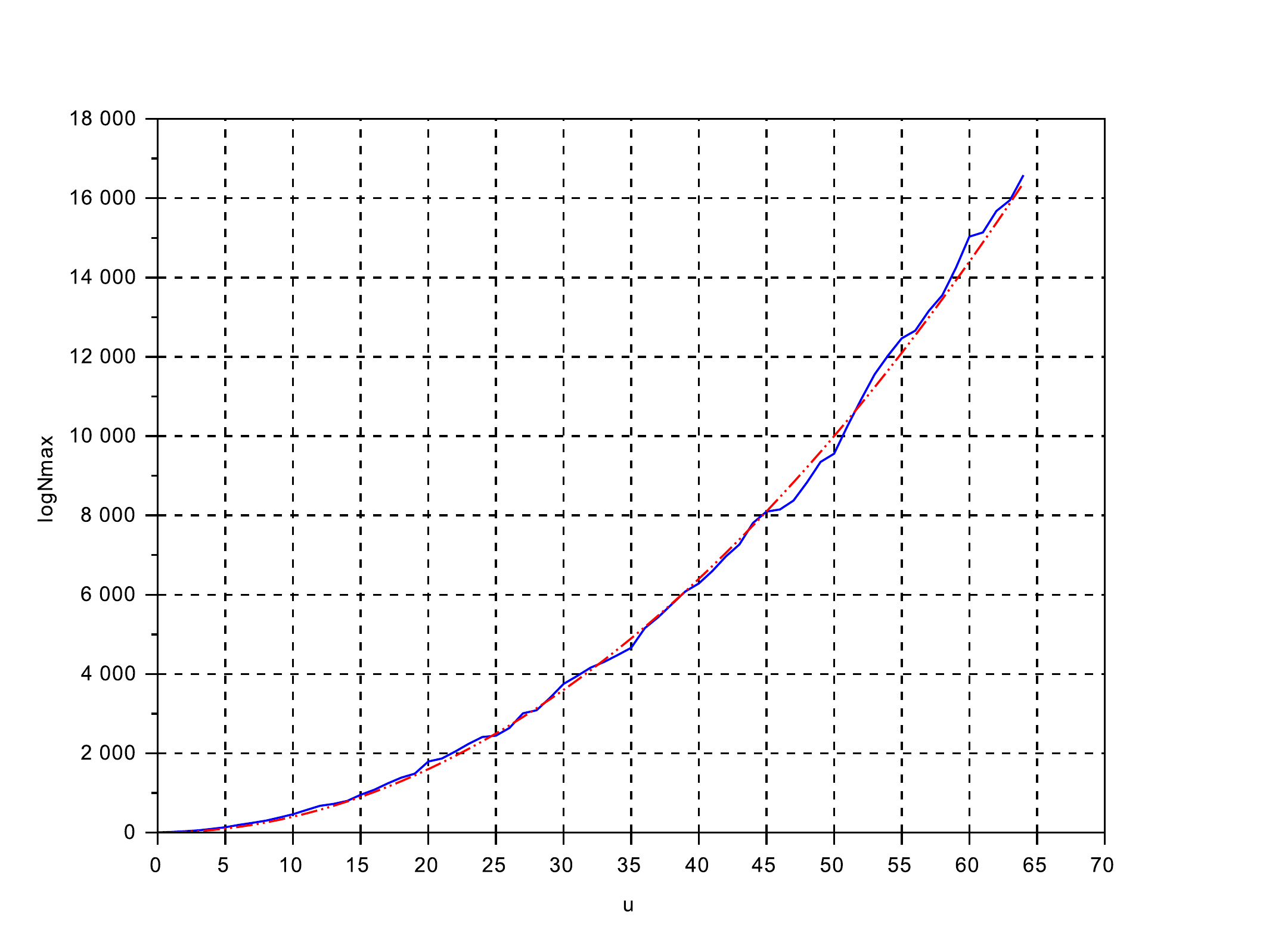}
	\caption{Values $\log N_{max}$ as a function of $u$ for primes $q_{23}$ versus function 
		$4\cdot u^2$ (dotted-dashed line). }
	\label{fig1}
\end{figure}
\begin{figure}
	\centering
		\includegraphics[height=0.4\textheight]{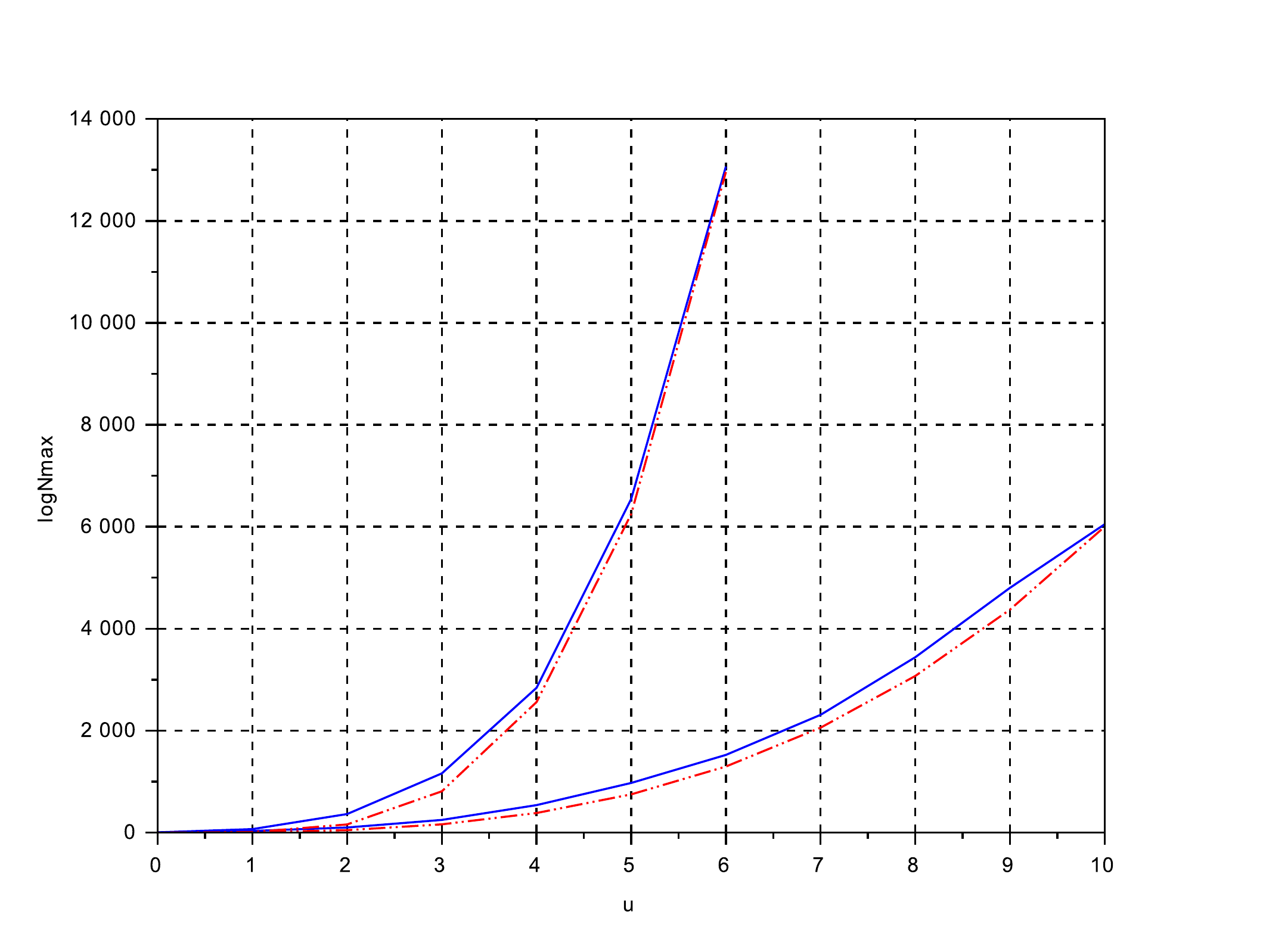}
	\caption{Values $\log N_{max}$ as a functions of $u$ for primes $q_{235}$ and $q_{2357}$ 
		compared to functions $6\cdot u^3$, and $10\cdot u^4$, 
		respectively (dotted-dashed lines).}
	\label{fig2}
\end{figure}

Let us introduce ordering of algorithms according to 
{\em rank} $u$. 
They are constructed using two types of small DFT modules: for sizes being prime 
numbers (\ref{primes23}):
\begin{equation}
	\label{order}
	q_{2p_2p_3\ldots } = 1 + 2^{s_1}{p_2}^{s_2}\ldots {p_d}^{s_d} 
	\quad {\rm for}\quad s_1 \le u+1, s_i \le u, i = 2,3,\ldots 
\end{equation}
where $p_2 = 3, p_3 = 5, \ldots $, and for sizes being 
powers of primes up to: $2^{u+3}$, and ${p_i}^{u+1}$ \cite{stas86}, section \ref{adds}. 
The only prime for $u=0$ is $3=2^1+1$, for $u=1$ the additional primes in class $q_{23}$ are $5=2^2+1$, 
$7=2^1\cdot 3^1 +1$, and $13=2^2\cdot 3^1+1$, in class $q_{235}$ they are complemented by 
$11=2^1\cdot 5^1+1$, $31=2^1\cdot 3^1\cdot 5^1+1$, $61=2^2\cdot 3^1\cdot 5^1+1$, and so on. 
In Figure \ref{map} new primes $q_{23}$ for $u = 10$ can be found. 
$M_{max}$ for rank $u>0$ is not greater than $\frac{2}{3}6^u$ for class $q_{23}$, $\frac{8}{15}30^u$ 
for $q_{235}$, and $\frac{48}{105}105^u$ for $q_{2357}$\footnote{For a $p$ prime there are 
	$(p-1)p^{r-1}$ natural numbers that are mutually prime with $p^r$, and smaller than $p^r$.}. 
$N_{max}$ is computed by 
multiplication of all numbers for all ranks\footnote{Except for divisors of powers of primes} up to $u$. 

Figures \ref{fig1} and \ref{fig2} show growth of maximum algorithm size $N_{max}$ as a function of $u$ for primes 
$q_{23}, q_{235}$, and $q_{2357}$, 
in fact, increase of number $\log_2 N_{max}$ is shown. 
The growth rates seem to be $O(u^2)$, $O(u^3)$, and $O(u^4)$ respectively, which
supports evaluations done in Appendix A. This means that size of   
$N_{max}$ is $O(2^{u^2})$, $O(2^{u^3})$, and $O(2^{u^4})$. 
Let us assume that polynomial products are computed using algoritm 1F. Then, multiplicative 
complexity of algorithms for primes $q_{23}$ is $O(N_{max}\log M_{max}) = O(2^{u^2}\cdot u)$, while $u$ is 
the square root of $u^2$. Similarly, $u$ is the cubic root of $u^3$, and fourth degree root 
of $u^4$. This is the last step of proof of the following Theorem (for whole proof see Appendix A):
\begin{quote}
	{\bf Theorem 3:} If polynomial products 
	are computed using 
	Algorithm 1F, then there exist fast DFT algorithms having multiplicative complexity
	\[
		O(N \log ^{1/m} N), 
	\]	
	i.e. lower than their arithmetical complexity $O(N\log N)$, $m$ is the class index (\ref{primes23}).
	
\end{quote}
In proof of Theorem 2 it is assumed that polynomial products are computed using Algorithm 2 
by DFT algorithms having the same multiplicative complexity as the main one, i.e. $O(N\log^c \log N)$.

Numbers of primes 
for algorithms of rank $u$ for primes $q_{23}$, $q_{235}$, $q_{2357}$ 
are $O(u)$, 
$O(u^2)$, and $O(u^3)$, respectively, which is in concordance with results from Appendix A, too. 
For example, they are usually 4 new primes $q_{23}$ when going from rank $u-1$ to $u$, 
hence, approximately $4u$ primes for rank $u$. The result can be checked by counting primes in Figure \ref{map}.

Let us consider multiplicative complexity of DFT algorithm for the greatest number $u=64$ in 
Figure \ref{fig1}, $N_{max}= 2^{16582.2}$. 
Considerations are simplified when parameters $\mu (N)$ and $a_M$ are introduced:
\begin{equation}
	\label{mu}
	\mu = \mu (N) = \frac{M(N)}{N}, \quad  a_M = \frac{M(N)}{N\log_2 N}.
\end{equation}
For the best radix-2 FFTs 
$\mu (N)$ 
is $\mu (N_{max})=\log N_{max} - 3 = 16579.2$ (\ref{fftma}). 
The greatest block of {\sf H} in (\ref{EHD}) is of size 
$M_{max}=2^{65}3^{63}$ (rank of cyclotomic polynomial equivalent to $P_{2^{65}3^{64}}(Z)$), 
which is approximately $2^{165}$. If in Algorithm 1F one of forward and inverse reduced split-radix FFTs 
from \cite{stas22} are used, then 
$\mu (N_{max})$ is upper bounded by  
$2\cdot \mu (2^{165})+3=2\cdot a_M\cdot 165+3\le 294$, see the next paragraph, 
for comparison, $\log_2 ^{1/2}N_{max}=128.8$, Theorem 3. 
The polynomial products can be computed also by Algorithm 2.  
We need here a forward and inverse DFT algorithm of size $2\cdot M_{max}$, or little more, 
that is one of rank $u=6$, see Fig.\ref{fig1}, which in the worst case 
have polynomial products of maximum size $2^7\cdot 3^5=31104$. 
They can be computed in the same way by the 65520-point algorithm.   
Number of multiplications for the last algorithm is 262820, Table I, hence, 
$\mu (65520) = 262820/65520 \approx 4.01$. Then, the worst case upper bound on $\mu (N)$ 
for a rank $u=6$ DFT algorithm is 
$\mu_2=\frac{65520}{31104}\cdot (2\cdot \mu(65520) +3)\approx 23.2$. 
Finally, 
assume that the size of rank $u=6$ DFT algorithm is $(2+\epsilon)\cdot M_{max}$, which leads to 
upper bound on 
$\mu (N_{max})=(2+\epsilon)(2\cdot \mu_2 +3)$, 
somewhat above 100, compare with $\log_2 \log_2 N_{max} = 14.02$, Theorem 2. 

It is interesting how Algorithm 2 compares to Algorithm 1F for exact data, i.e. 
what is the cost of computing 31104-point polynomial product by 
Algorithm 1F. Formulae from \cite{stas94} show that reduced 93312-point FFT 
requires 406350 multiplications (FFT part for DFT indices mutually prime with $N$). This leads to 
$\mu_{1F}=2\cdot \frac{406350}{31104}+3\approx 29.1>\mu_2$. 
This is as expected, nevertheless, Algorithm 1F requires less additions. 
Here the $a_M$ coefficient is $a_M 
\approx 0.875$, it is 
used in evaluation of $\mu (N_{max})$ in the previous paragraph. 
Taking into account that for 
"parental" 93312-point FFT $a_M \approx 0.73$ that it does not increase with $N$ \cite{stas22}, 
and that the difference in $a_M$ values between full FFT, 
and its reduced part diminish with growing $N$, 
actual bound is probably even smaller.

\section{Conclusion}
One  of  open  questions  posed  in   \cite{aho}   is,   if   the 
arithmetical complexity  of  DFT  and  hence  polynomial  algebra 
algorithms is $O(N\log N)$, or it is smaller. This paper  provides  an 
interesting comment to this question --- the  DFT  can  be  computed 
using $O(N\log N)$ arithmetical operations, among which  all  but  few 
are  additions   and   subtractions.   Namely,   the   number   of 
multiplications needed is $O(N\log ^{c} \log N)$ for $c\le 1$. 
The algorithms are constructed using nesting 
technique guaranteeing 
that arithmetical complexity
of the algorithms is $O(N\log N)$. 
The method is based on extensive use 
of polynomial transforms, which implies special choice  of component DFT modules. The 
limits on algorithm multiplicative complexity are obtained for an assumption that the density of 
primes being sizes of small DFT modules in the set of potential DFT sizes 
can be approximated by the same relations as for 
primes in general. Experimental results confirm that this premise is reasonable. What is important, 
for practical range of sizes the new algorithms have approximately the same   
arithmetical complexities as the best known FFTs, while require much less multiplications.





\appendix 

\section{Proofs}
\label{proof}

Let us begin with evaluation of lower bound on ''density'' of prime numbers from the class 
$q_{23}$ (\ref{primes23}). 
\begin{quote}
	{\bf Corollary 1:} For a natural number $s_{max}$ there exist $O(s_{max})$ 
	natural numbers $s_2 < s_{max}$ for which exists at least one prime 
	number of the form $q_{23}$ (\ref{primes23}).
\end{quote}
{\em Discussion instead of proof:} In book \cite{riben} discussion on existence of primes 
$k2^n+1$ for each $k$ and some $n$ is reported, $k, n$ are natural numbers. The conclusion
is that $k$, for which $k2^n+1$  
numbers are composite for all natural $n$, are extremely rare. The only 
known ones are 
Sierpi\'{n}ski numbers. They meet some congruences linked with divisors of fifth 
Fermat number, 
hence, for them usually $k\ne 3^{s_2}$.
It is then highly improbable that other than Sierpi\'{n}ski numbers never being primes 
are for almost all $k$ of the form $k=3^{s_2}$, of course, if such numbers exist at all.
{\em Q.E.D.}
\begin{quote}
	{\bf Theorem 4:} With possible exception of Fermat prime numbers the number of primes of the form 
	(\ref{primes23}) is infinite.
\end{quote}
{\em Proof:} With $s_{max} \rightarrow \infty $, $O(s_{max}) \rightarrow \infty $, 
too.  Number classes $q_{235\ldots }$ contain $q_{23}$ one {\em Q.E.D.}

We are now (almost) sure that asymptotic considerations have sense for this 
algorithm class. 

In the following derivations it is assumed without proof 
that numbers (\ref{primes23}) 
are not peculiar, i.e. that 
density of primes among them can be 
evaluated using the same approximations as for randomly chosen natural numbers. We 
need a condition stronger than Corollary 1:
\begin{quote}
	{\bf Corollary 2:} For natural numbers $s_2 \le s_{max}$, and  
	$s_1 \le s_{max}'$,  $s_{max}' \approx s_{max}$ there exist $O(s_{max})$ 
	prime numbers from the class $q_{23}$ (\ref{primes23}).  
\end{quote}

{\em Proof:} Let us consider average amounts of prime numbers in the following series, 
see Figure \ref{map}, new rank $u$ primes, section \ref{mults}:
\[
p_{3i} = 2^{u+1} 3^i + 1, \mbox{    } i=0,1,\ldots , u; 
\]
\[
p_{2i} = 2^i 3^{u} + 1, \mbox{    } i=1,\ldots , u + 1; 
\]
The average 
distance between a prime $p$ and the following one is $O(\log p)$ \cite{riben}, \cite{10}, 
hence, we can evaluate the amount of primes as follows:
\begin{equation}
	\label{pcount}
	p_{count} \approx \sum _{i} \frac{1}{\log p_{ji}}, \mbox{    } j=2,3.
\end{equation}
The denominator in (\ref{pcount}) varies in small range of values, 
e.g. for $p_{3i}$ its range is from $(u+1)\log 2$ to 
$(u+1)\log 2 + c_2 (u+1)\log 3$, $c_2$ is a constant, hence, (\ref{pcount}) can be 
bounded by, $v = u + 1$:
\begin{equation}
	\label{pridens}
	\frac{c_2 v}{v\log 2} = \frac{c_2}{\log 2} 
	> p_{count} 
	> \frac{c_2 v}{v\log 2 + c_2 v\log 3} 
	= \frac{c_2 }{\log (2\cdot 3^{c_2})} = O(1),
\end{equation}
similar evaluation is true for series $p_{2i}$. Then, taking all 
primes for $u = 1, 2, \ldots , s_{max}$ we obtain $O(s_{max})$ primes
{\em Q.E.D.}

\begin{quote}
	{\bf Corollary 3:} Multiplicative complexity of DFT algorithms of size (\ref{anyn})
	for primes from class $q_{23}$ 
	(\ref{primes23}) is 
	\[
	M(N) = 
	O(N\log \log N)
	\]
\end{quote}
{\em Proof:} Let us consider primes from the set $q_{23}$ (\ref{primes23}), $i$ is the prime index:
\begin{equation}
	\label{q23}
	q_{i} = 2^{s_{1i}} 3^{s_{2i}} + 1 
	\quad {\rm where} \quad s_{1i} \le u + 1, 
	\quad s_{2i} \le u
\end{equation}
For such primes 
$M_{max}$ is, see first footnote in section \ref{mults}:
\begin{equation}
	\label{maxM}
	M_{max} 
	\le \frac{1}{3} 2^{
		u+1} 3^{u} 
	= 3^{c_3 u}
\end{equation}
$c_3$ is a constant. 
On the other hand, the maximum size of the DFT is\footnote{Inclusion of Rader-Winograd DFTs for 
	$2^{k_1}$, $3^{k_2}$ does not change the final evaluation.}
(\ref{anyn}), (\ref{q23}):
\[
N_{max} = 
\prod _{i} q_i 
\]
for all $i$. Then:
\[
N_{max} \ge 
\prod _{i} (2^{s_{1i}} 3^i + 1) >
\prod _{i} 2^{s_{1}} 3^i \ge 
2^{2c_4 u} 3^{c_4 u^2} 
\]
\begin{equation}
	\label{Nmax}
	N_{max} > 3^{c_4 u^2} ,
\end{equation}
in the derivation it is assumed that for each power of 3 only one prime\footnote{
	Actually there are usually 4 primes in the series, section \ref{mults} .}
is taken into 
consideration (i.e. $s_{2i} =i$), $i$ takes almost all values between 0 and $u$, i.e. 
there are not more than $2c_4 u$ primes (values of $i$), and that $s_1 \ge 1$ 
but is taken small enough to fulfill the inequality in the center, $c_4$ is a constant. 

A polynomial product 
can be computed using Algorithm 2,  
based on the DFT algorithm derived here, but of much smaller size, for $\mu (N)$ see (\ref{mu}) :   
\begin{equation}
	\label{mucc}
	\mu (M\mbox{-point PP}) \approx c_1 + 2 \mu(2M\mbox{-point DFT}) 
	\approx 2 \mu(2M\mbox{-point DFT}).
\end{equation}
PP - polynomial product, $c_1$ is a constant. 
Then, we are seeking for a function expressing $\mu (N)$ for which (\ref{maxM}), (\ref{Nmax}):
\[
\mu(3^{c_4 u^2}) \approx 2 \mu (3^{c_3 u}),
\]
which appears to be $\log \log _b (N)$, $b=3^{c_3}$, at least for $\log u >> \log \frac{c_4}{c_3}$. 
Hence:
\[
M(N_{max}) = O(N_{max} \log \log N_{max}).
\]
{\em Q.E.D.}

Assume now that polynomial products of size (\ref{maxM}) are computed using Algorithm 1F requiring 
$O(M_{max} \log M_{max})$ multiplications, i.e.:
\[
M(M_{max}) \le c_5 3^{c_3 u} \log 3^{c_3 u} \le c_6 u M_{max}
\]
$c_5, c_6$ are constants. 
As (\ref{Nmax}):
\[ 
u < \sqrt {\frac{1}{c_4} \log _{ 3} N_{max}},
\]
then
\begin{equation}
	\label{simplified}
	M(N_{prod}) = O(N_{max} \log ^{c_l} N_{max}), \mbox{    } c_l = 1/2 < 1.
\end{equation}
DFT algorithms implemented in Algorithm 2 and 1F are different, nevertheless the used in 
Algorithm 1F radix-6 or similar FFT are very efficient \cite{stas22}. This means that  
Algorithm 2 requires $O(M_{max}\log M_{max})$ extra additions, as it uses 
FFT algorithm of size $2 M_{max}$. 
As 
\[
M_{max}\log M_{max}/ N_{max}\log N_{max} \rightarrow 0
\]
for growing $N_{max}$, the increase in number of additions is negligible, and conditions 
(\ref{M2A}), (\ref{Acond}) are met also for the algorithm from Corollary 3.

We are now ready to prove Theorems 2 and 3:\newline
{\em Sketch of Proof:} 
Let us consider series of numbers from class $q_{235}$ 
(\ref{primes23}):
\begin{equation}
	\label{cons235}
	q_{ijk} = 2^{s_{1i}} 3^{s_{2j}} 5^{s_{3k}} + 1 
	\enspace {\rm where} \enspace s_{1i} \le u+1, \enspace s_{2j}, s_{3k} \le u
\end{equation}
Similarly as in proof of Corollary 2, we want to count primes in the series, 
here they are two-dimensional ones for indices 
$2^{u+1} 3^j 5^k$, $2^{i} 3^{u} 5^k$, $2^{i} 3^j 5^{u}$, $i = 1, 2, \ldots u+1$; $j, k = 0, 1, \ldots u$.
As previously, the numbers of primes in one-dimensional series 
is $O(1)$ (\ref{pridens}), hence, $O(u)$ for two-dimensional ones. When repeating reasoning from proof of Corollary 3, 
$M_{max}$ is:
\[
M_{max} = \frac{4}{15} 2^{
	u+1} 15^{u}  < \frac{1}{2} 15^{c_8 u},
\]
while the maximum DFT size is:
\[
N_{max} > \prod _{j} \prod _{k} 2^{s_1} 3^j 5^k > 2^{s_1 u^2} 15^{c_7 u^3} > 15^{c_7 u^3},
\]
$c_7 $ is a constant. As previously, it is assumed that there is only one prime
in an one-dimensional series,
and that $s_1 \ge 1$, but 
small enough to fulfill the central inequality, $c_7$ has the same role as $c_4$. For Algorithm 2 we need a function for which:
\[
\mu (15^{c_7 u^3}) \approx 2 \mu (15^{c_8 u}),
\]
and this is $\log ^c \log _b (.)$ for $c=\log_3 2 < 1$, $b=15^{c_7}$. 
Note that when $m$ $p_i $ primes are used in (\ref{cons235}), 
then $N_{max}=O(2^{u^m})$, while $M_{max}=O(2^u)$, and 
multiplicative complexities of the algorithms are described by the same functions, 
only the constant $c $ diminishes.

As observed in section \ref{mults}, in Theorem 3 
the constant $c_l=1/m$ (\ref{simplified}), if class $m$ of primes from (\ref{primes23}) is  
considered. Addition counts are estimated in 
section \ref{adds}. {\em Q.E.D.}


\newpage

\section{Few real algorithms} 
\label{reala}

In Table I 
data for some interesting algorithms for smallest primes from classes $q_2$ and $q_{23}$ 
can be found, $u = 0, 1$. 
Trivial multiplications are not counted. 
The latter family is constructed using widely known DFT 
modules \cite{nuss81}, 
plus the 13-point one. 
Numbers of arithmetical operations 
are compared to those obtained from formulae (rounded up):
\begin{equation}
	\label{fftma}
	\begin{array}{l}
		\mbox{mfft }(N) = N\log _{2}N - 3 N + 4,  \\
		\mbox{afft }(N) = 2 N\log _{2}N + \mbox{mfft }(N),
	\end{array}
\end{equation}
which for $N$ being power of 2 describe multiplication and addition counts for the split-radix 
FFT \cite{srfft} (3 multiplications and 3 additions per complex multiplication). As can be seen, 
majority of algorithms in Table B.1 have {\em smaller} arithmetical complexities than that for 
the split-radix FFT formula. The difference in percents of the number of arithmetical operations is 
provided in column ''saldo''. It should be noted that improvement in arithmetical complexity given by 
algorithm from \cite{optfft} being $-5.56\%$ is asymptotic, up to $N=16$ numbers of operations are identical to 
those for the split-radix FFT, then the numbers gradually diverge. 
Of course, numbers of multiplications for algorithms from Table B.1 are 
much smaller than for any FFT. Many of the algorithms are also better than any presented 
in the bibliography for the same DFT size \cite{nuss81}, even if some WFTA improvements are done, 
see e.g. \cite{burrb}. 

Note that grouped below a line Fermat primes based algorithms require more 
multiplications, but less additions than those for $q_{23}$, nevertheless, if the sum of 
arithmetical operations is considered, they are very interesting, too. In particular, 
impressive is the performance of the 8160-point algorithm (without FFTs, first line), especially that it is 
constructed using such ''ineffective'' Rader-Winograd DFT modules as the 17-point and 32-point 
ones\footnote{The polynomial product modulo $P_{16}(Z) = Z^8 + 1$ in 17- and 64-point DFTs is computed using 27 
	multiplications and 57 additions \cite{nuss81}.}.

In Fermat prime based algorithms with "FFT modules" for $N=8160$, and 16320 
(second lines) some of 32- or 64-point DFT modules are FFTs, not Rader-Winograd algorithms. Multipliers of these 
FFTs are nested (\ref{EHD}), prescription how it is done can be found in \cite{stas87}.

The computed on the basis of Theorem 1 theoretical minimum on the number of multiplications for 65520-point 
DFT is 217556 (108778 non-trivial real or imaginary multipliers, 
$2N = 131040$). This is only $17.2\% $ less than the number from 
Table B.1. The greatest DFT algorithm in Table B.1 that meets the theoretical lower limit on the number of 
multiplications is for $N = 504$.

\onecolumn
\begin{table}
	
	\begin{center}
		
		\caption{Some real almost multiplierless DFT algorithms \cite{stas87}.} 
		{\small
			\begin{tabular}{||r|c|r|r|r|r|r||} \hline\hline
				$N$   & $N$ divisors & mults &  adds  &  mfft &  afft  & saldo \\  \hline
				24  & $8\times 3$          &    36 &    252 &     42 &    262 & $-5.26\% $ \\
				48  & $16\times 3$         &    92 &    636 &    128 &    664 & $-8.08\% $ \\
				120  & $8\times 3\times 5$  &   276 &   2076 &    473 &   2130 & $-9.64\% $ \\
				240  & $16\times 3\times 5$ &   596 &   4836 &   1182 &   4977 & $-11.80\% $ \\
				504  & $8\times 9\times 7$  &  1380 &  13532 &   3017 &  12066 & $-1.13\% $ \\
				840  & $8\times 3\times 5\times 7$ & 2580 & 23460 & 5644 & 21964 & $-5.68\% $ \\
				1008  & $16\times 9\times 7$ &  3116 &  30124 &   7037 &  27151 & $-2.77\% $ \\
				1680  & $16\times 3\times 5\times 7$ & 5492 & 51924 & 12964 & 48964 & $-7.29\% $ \\
				2520  & $8\times 9\times 5\times 7$ & 8340 & 85948 & 20918 & 77866 & $-4.55\% $ \\
				5040  & $16\times 9\times 5\times 7$ & 17732 & 187124 & 46872 & 170848 & $-5.91\% $ \\
				6552  & $8\times 9\times 7\times 13$ & 23460 & 276812 & 63412 & 229541 & $+2.50\% $ \\
				10920 & $8\times 3\times 5\times 7\times 13$ & 41028 & 471156 & 113732 & 406709 & $-1.59\% $ \\
				13104 & $16\times 9\times 7\times 13$ & 49484 & 592972 & 139925 & 498391 & $+0.65\% $ \\
				21840 & $16\times 3\times 5\times 7\times 13$ & 84620 & 1007796 & 249301 & 878934 & $-3.17\% $ \\
				32760 & $8\times 9\times 5\times 7\times 13$ & 127572 & 1620076 & 393112 & 1375889 & $-1.21\% $ \\
				65520 & $16\times 9\times 5\times 7\times 13$ & 262820 & 3436820 & 851741 & 2948335 & $-2.64\% $ \\ \hline
				2040  & $8\times 3\times 5\times 17$ & 10212 & 68844 & 16312 & 61169 & $+2.03\% $ \\
				4080  & $16\times 3\times 5\times 17$ & 20540 & 151668 & 36701 & 134575 & $+0.54\% $ \\
				8160  & $32\times 3\times 5\times 17$ & 42208 & 327264 & 81558 & 293626 & $-1.52\% $ \\
				&  FFT modules              & 41908 & 326364 &        &       & $-1.84\% $ \\
				16320 & $64\times 3\times 5\times 17$ & 91416 & 719904 & 179432 & 636208 & $-0.53\% $ \\ 
				&  FFT modules              & 88716 & 711804 &        &       & $-1.85\% $ \\ \hline\hline
				
			\end{tabular}
		}
	\end{center}
\label{exalg}	
\end{table}


\begin{thebibliography}{00}
	
		\bibitem{fft65} J.W. Cooley, J.W. Tukey, ''An algorithm for machine computation of complex 
	Fourier series,'' Math. Comput., vol. 19, pp. 297--301, 1965.
	
	\bibitem{au6769} ---, ''Special issue on fast Fourier transform,'' IEEE Trans. Audio Electroacoust., 
	vol. AU-15, and vol. AU-17, June 1967, and June 1969.
	
	\bibitem{aho}   A.V. Aho, J.E. Hopcroft, J.D. Ullman,  ``The  design  and 
	analysis of computer algorithms'', Addison-Wesley, 1974.
	
	\bibitem{wino80}
	S. Winograd, ''On the multiplicative complexity of the discrete Fourier Transform'', 
	Adv. Math., vol. 32, 83--117, 1979.
	
	\bibitem{wino} 
	S. Winograd, ``On computing the discrete Fourier transform'', Proc. Nat.
	Acad. Sci. U.S.A., vol. 73, pp. 1005--1006, 1976.
	
	
	\bibitem{silver}
	H.F. Silverman, ``An introduction to programming the Winograd Fourier
	transform algorithms (WFTA)'', IEEE Trans. Acoust., Speech, Signal Proces.,
	vol. ASSP-25, pp. 152--165, 1977.
	
	\bibitem{burrb}
	C.S. Burrus, T.W. Parks, ``DFT/FFT and convolution algorithms'', John Wiley,
	New York, 1985.
	
	
	\bibitem{nuss}
	H. J. Nussbaumer, ``New algorithms for convolution and DFT based on
	polynomial transforms'', Proc. ICASSP-78, pp. 638--641, 1978.
	
	\bibitem{nussq}
	H.J. Nussbaumer, P. Quandalle, ``Fast computation of fast Fourier
	transform using polynomial transforms'', IEEE Trans. Acoust., Speech,
	Signal Proces., vol. ASSP-27, pp. 169--181, 1979.

	
	\bibitem{srfft}
	P. Duhamel, H. Hollmann, ``Split-radix FFT algorithm'', Electron. Lett.,
	vol. 20, No. 1, pp. 14--16, 1984.	
	

	\bibitem{optfft}
	S.G. Johnson, M. Frigo, "A modified split-radix FFT with fewer arithmetic 
	operations", IEEE Trans. Signal Processing, vol. 55 (1), pp. 111–119, 2007.
	
	\bibitem{stas22}
	R. Stasinski, "Split multiple radix FFT", Proc. EUSIPCO 2022, pp. 2251-2255, 2022.
	
	\bibitem{martens}
	J.-B. Martens, ”Recursive cyclotomic factorization - a new algorithm
	for calculating the Discrete Fourier Transform”, IEEE Trans. Acoust.,
	Speech, Signal Proces., vol. ASSP-32, pp. 750–761, 1984.
	
	\bibitem{stas94}
	R. Stasinski, ”Radix-K FFT using K-point convolutions”, IEEE Trans.
	Signal Proces., vol. 42, pp. 743–750, 1994.
	
	\bibitem{harvey}
	D. Harvey, J. van der Hoeven, G. Lecerf, "Even faster integer multiplication", 
	J. Complexity, vol.36, pp. 1–-30, 2016.  
	
	\bibitem{harvey19}
	D. Harvey, J. van der Hoeven. ``Integer multiplication in time O(n log n)'',  
	hal-02070778, 2019.
	
	\bibitem{filterdft}
	L. Bluestein, "A linear filtering approach to the computation of discrete fourier transform", Audio and Electroacoustics, IEEE Transactions on, vol. 18, no. 4, pp. 451–455, 1970. 
	
	\bibitem{11} C.M. Rader, ``Discrete Fourier transform when the  number 
	of data samples is prime'', Proc. IEEE,  Vol.  56, pp. 1107--1108, 1968.
	
	\bibitem{stas87} R. Stasinski,  ``Prime  factor  DFT  algorithms  for  new 
	small-N DFT modules'', IEE Proc., Pt. G, Vol. 134, pp. 117--126, 1987.
	
	\bibitem{nuss81}
	H.J.  Nussbaumer,  ``Fast Fourier transform  and  convolution 
	algorithms'', Springer-Verlag, 1981.
	
	\bibitem{stas86}  R.  Stasinski,  ``Easy  generation  of  small-N  discrete 
	Fourier transform algorithms'', IEE Proc.,  Pt.  G,  Vol. 133, pp. 133--139, 1986.
	
	\bibitem{duh84} 
	P. Duhamel, H. Hollmann, "Existence of a $2^n $ FFT algorithm with a number of 
	multiplications lower than $2^{n+1}$'', Electron. Lett., vol. 20, No. 17, 
	pp. 690--692, 1984.
	
	\bibitem{stas92}
	R. Stasi\'{n}ski, ``Reducing multiplicative complexity of polynomial
	algebra algorithms'', Proc. IFIP'92 Congress, pp. 268--274, 1992.
	
	
	
	\bibitem{riben} P. Ribenboim, ''The little book of big primes,'' Springer-Verlag, 1991, 1996.
	
	
	
	
	
	\bibitem{10}  K. Ireland,  M.  Rosen,  ``A  classical  introduction  to 
	modern number theory'', Springer-Verlag, 1982.
	
	
	

	
	
	
	
	
\end{thebibliography}
\end{document}